# Robust thalamic nuclei segmentation from T1-weighted MRI

Julie P. Vidal[1,2], Lola Danet[2,3], Patrice Péran[2], Jérémie Pariente[2,3], Meritxell Bach Cuadra[4], Natalie M. Zahr[5,6], Emmanuel J. Barbeau[1], and Manojkumar Saranathan*[7]

[1]CNRS, CerCo (Centre de Recherche Cerveau et Cognition) - Université Paul Sabatier, Toulouse, France, [2]INSERM, ToNiC (Toulouse NeuroImaging Center) - Université Paul Sabatier, Toulouse, France, [3]Hôpital Purpan, Centre Hospitalier Universitaire de Toulouse, Département de Neurologie, Toulouse, France, [4]FBM (Faculté de Biologie et de Médecine) - Université de Lausanne, Lausanne, Switzerland, [5]Center for Health Sciences, SRI International, Menlo Park, California, USA, [6]Department of Psychiatry & Behavioral Sciences - Stanford University School of Medicine, Stanford, California, USA, [7]Department of Radiology - University of Massachusetts Chan Medical School, Worcester, Massachusetts, USA

* Email: *Manojkumar.Saranathan@umassmed.edu*

# Abstract

Accurate segmentation of thalamic nuclei, crucial for understanding their role in healthy cognition and in pathologies, is challenging to achieve on standard T1-weighted (T1w) magnetic resonance imaging (MRI) due to poor image contrast. White-matter-nulled (WMn) MRI sequences improve intrathalamic contrast but are not part of clinical protocols or extant databases. Here, we introduce Histogram-based polynomial synthesis (HIPS), a fast preprocessing step that synthesizes WMn-like image contrast from standard T1w MRI using a polynomial approximation. HIPS was incorporated into our Thalamus Optimized Multi-Atlas Segmentation (THOMAS) pipeline, developed and optimized for WMn MRI. HIPS-THOMAS was compared to a convolutional neural network (CNN)-based segmentation method and THOMAS modified for T1w images (T1w-THOMAS). The robustness and accuracy of the three methods were tested across different image contrasts, scanner manufacturers, and field strength. HIPS-synthesized images improved intra-thalamic contrast and thalamic boundaries, and their segmentations yielded significantly better mean Dice, lower percentage of volume error, and lower standard deviations compared to both the CNN method and T1w-THOMAS. Finally, using THOMAS, HIPS-synthesized images were as effective as WMn images for identifying thalamic nuclei atrophy in alcohol use disorders subjects relative to healthy controls, with a higher area under the ROC curve compared to T1w-THOMAS (0.79 vs 0.73).

1. Introduction

The thalamus is a deep brain structure on either side of the third ventricle, comprised of multiple nuclei [1]. Physiologically, these nuclei are often classified as first order nuclei (e.g. lateral and medial geniculate nuclei), relaying sensory or motor information to the cortex, higher order nuclei (e.g. pulvinar) involved in cognition through cortico-thalamo-cortical circuits [2] and the thalamic reticular nucleus. Several neurodegenerative, neurological, and neuropsychiatric conditions involve the thalamic nuclei such as alcohol use disorder [3], schizophrenia [4], Alzheimer's disease [5], chronic pain syndrome [6], epilepsy [7] and stroke [8]. Thus, visualization and characterization of thalamic nuclei are crucial in understanding their function as well as their relationship in the healthy brain or in pathological conditions. This need is even more critical for pathologies for which clinical care is based on accurate targeting of specific nuclei, such as deep brain stimulation (DBS) for the treatment of essential tremor [9] [10], chronic pain syndrome [11] or drug-resistant epilepsy [12]. Other therapeutic approaches targeting specific thalamic nuclei include Magnetic Resonance Guided Focused Ultrasound thalamotomy (MRgFUS) [13], gamma knife surgery [14], radio frequency surgery [15], and microsurgical resection [16].

Segmentation of thalamic nuclei is challenging because of their poor contrast on standard T1 and T2 weighted (T1w, T2w) MRI, the most popular sequences in routine neuroimaging protocols. To partly address this issue, atlases such as the Schaltenbrand-Warren and the Morel atlas [17] [1] have been developed. These atlases have been used for manual delineations of nuclei on MRI data as well as in neurosurgical targeting using stereotactic coordinates [18]. Manual delineation of thalamic nuclei requires special expertise, is time consuming, and is therefore not ideal for analysis of large datasets. In addition, the small size of several nuclei makes them difficult to delineate. A three-dimensional reconstruction of the Morel Atlas in MNI space has been developed to help automate the segmentation of thalamic nuclei [19]. However, such atlas-based approaches ignore important inter-individual and inter-thalamic variability of the size, shape and location of thalamic nuclei, leading to compromised accuracy.

To address the issue of automated thalamic nuclei parcellation at a subject-level and exploit image contrast, several approaches have been explored. One of the earliest methods, proposed by Behrens et al. [20], used probabilistic tractography from Diffusion Tensor Imaging (DTI) data, mapping structural connectivity between different thalamic regions and specific cortical regions based on white matter anisotropy (e.g. mediodorsal nucleus to prefrontal cortex, lateral geniculate nucleus to visual cortex and so on), resulting in 6 thalamic regions corresponding to the 6 seed regions. Other diffusion MRI based methods have used local information from the diffusion tensor at a voxel level to parcellate the thalamus. Mang et al. [21] used k-means clustering of the dominant diffusion orientation while Battistella et al. [22] used k-means clustering of the spherical harmonic coefficients of the orientation distribution function to parcellate the thalamus into 6 regions. Since the thalamus is mainly composed of isotropic grey matter, the direction information computed from the diffusion tensor tends to be noisy. Further, the spatial resolution limitations of the underlying echo planar imaging acquisition results in a small number of clusters rather than precise, anatomically defined nuclei. Functional MRI based segmentation approaches have also been proposed using the idea of functional connectivity to cortical ROIs employing seed-based [23], or Independent Components Analysis (ICA) [24]. These also resulted in 6 thalamic regions corresponding to structural connectivity results of Behrens et al. In contrast, to parcellate the thalamus into 15 clusters, Kumar et al. [25] proposed an ICA based functional parcellation while Van Oort et al. [26] used time courses of instantaneous connectivity. Other methods based on susceptibility weighted imaging or quantitative susceptibility mapping [27] [28] have typically relied upon manual segmentation and have been limited to targeting the VIM nucleus for DBS treatment of movement disorders.

Images from structural (anatomic) MRI methods such as T1w Magnetization Prepared Rapid Gradient Echo (MPRAGE) have high spatial resolution (typically 1mm isotropic), minimal distortion, and are commonly used for cortical segmentation but rarely used for thalamic nuclei segmentation due to poor intra-thalamic nuclear contrast. Iglesias et al, developed a probabilistic atlas that combined manual delineations from in vivo T1w MPRAGE data and ex-vivo histological data [29]. The Bayesian segmentation algorithm that uses this atlas to segment T1w MRI is part of the Freesurfer package. Variants of MPRAGE such as white-matter-nulled (WMn) [30], and grey-matter-nulled (GMn) [31] MPRAGE imaging significantly improves intra-thalamic contrast and permits better delineation of thalamic nuclei. We then developed Thalamus Optimized Multi-Atlas Segmentation (THOMAS) [32] that uses 20 WMn-MPRAGE prior datasets acquired at 7T and segmented manually using the Morel atlas as a guide combined with a joint-label fusion algorithm for thalamic nuclei parcellation of WMn-MPRAGE data. This method (WMn-THOMAS) divided the thalamus into 11 nuclei per hemisphere and was validated against manual segmentation.

While WMn-THOMAS is promising and has been used in several studies examining the role of thalamic nuclei in alcohol use disorder and multiple sclerosis [33] [34], WMn-MPRAGE sequences are neither part of commonly used clinical protocols nor available in existing data repositories like ADNI or OASIS. To segment conventional MPRAGE T1w data, THOMAS was recently modified to use mutual information (MI) instead of cross-correlation (CC) as the nonlinear registration metric [35] [36] and a majority voting algorithm for label fusion (T1w-THOMAS). While this method achieved good accuracy compared to WMn-MPRAGE for larger nuclei such as the mediodorsal or pulvinar, it was less accurate for segmentation of the smaller centromedian and habenular nuclei. One reason for suboptimal performance of the modified T1w-THOMAS method could be due to poor intrathalamic contrast and thalamic boundaries on standard T1w contrast images. To address this, a novel deep learning-based method [37] was proposed, that first synthesized WMn-MPRAGE-like images from T1w data and then segmented the synthetic WMn-like data, using two separately trained convolutional neural networks (CNN). This CNN-based method was shown to improve accuracy compared to direct segmentation of T1w data (using a different CNN trained directly on 3T T1w data). However, the adoption of this method has been limited due to the necessity of training the CNNs on new types of data (e.g. different field strengths or scanner manufacturer). Furthermore, this CNN training process is time consuming and often not possible due to lack of concordant T1w and WMn-MPRAGE data acquired concurrently on the same subjects.

Inspired by the success of the synthesis approach of Umapathy et al. [37] and leveraging WMn-THOMAS, we introduce here a new pre-processing step, Histogram-based Polynomial Synthesis (HIPS), to improve thalamic nuclei segmentation by synthesizing WMn-like images from T1w MRI using a polynomial function. To test the hypothesis that HIPS-THOMAS outperforms the T1w-THOMAS or CNN-based segmentation, quantitative performance metrics were used to assess performance across differing contrast (MPRAGE, SPGR, MP2RAGE), scanner manufacturers (Philips, GE, Siemens), and field strengths (3T, 7T). To assess clinical validity, results of HIPS-THOMAS segmentation volumes were compared to WMn-THOMAS results considering group differences in thalamic nuclei volumes between 21 individuals diagnosed with alcohol use disorders (AUD) and 23 healthy control subjects.

2. Methods

2.1 Histogram-based polynomial synthesis (HIPS)

WMn-like images were synthesized from T1w MRI using a training set of T1-MPRAGE and WMn-MPRAGE datasets acquired on the same subjects and fitting a polynomial function (SciPy, python) to a plot of T1-MPRAGE vs. WMn-MPRAGE intensity values. The two datasets were affine registered using Advanced Normalization Tools (ANTS) [38]. To render the fit scanner and subject independent, T1-MPRAGE images were normalized using the WM signal and WMn-MPRAGE images were normalized using the CSF signal, both extracted from their respective image histograms, prior to fitting. Histograms were computed using a cropped middle axial slice to focus on the bilateral thalami. To perform the normalization, the mode of the tissue of interest (WM in T1w, CSF in WMn-MPRAGE) was first computed and the highest value shared by at least 1% of voxels exceeding the mode was used for the normalization. This approach is similar to the WhiteStripe intensity normalization method [39], and is effective in removing the high intensity tail corresponding to artifacts and outlier intensities [40]. The function selected among those tested (linear, quadratic, cubic, and quartic) was the one that yielded the best segmentation results evaluated using three quantitative metrics- structural similarity index (SSIM), Dice coefficient and volume similarity index (VSI) [41].

2.2 The HIPS-THOMAS segmentation pipeline

HIPS-THOMAS is a variant of the THalamus Optimized Multi Atlas Segmentation (THOMAS) [32] method and is shown in Figure 1. The input T1w image is first cropped to cover both thalami which removes outliers from the skull/subcutaneous fat. Following the cropping step, HIPS preprocessing is applied, which first consists of a normalization step described earlier, ("NV", Fig. 1). The polynomial function is then applied followed by a contrast stretching step and finally images are rescaled to the highest WM value computed previously with the histograms to get standardized intensity ranges. It results in a WMn-like version of the cropped input T1w image. This HIPS processed cropped image is then nonlinearly registered ("Warp R") to a cropped average brain template from 20 WMn-MPRAGE priors using the cross correlation metric ("CC" in blue, Fig. 1). This nonlinear warp is inverted ("$R^{-1}$", Fig.1) and combined with the 20 precomputed warps from priors to the template ("$W_{piT}$", Fig. 1) to put the 20 manual segmentation labels in the input space. These 20 labels were then combined using a joint label fusion algorithm to generate the output's final parcellation. The segmented thalamic nuclei include Anteroventral (AV), Ventral anterior (VA), Ventral lateral anterior (VLa), Ventral lateral posterior (VLp), Ventral posterolateral (VPL), Pulvinar (Pul), Lateral geniculate (LGN), Medial geniculate (MGN), Centromedian (CM), Mediodorsal-Parafascicular (MD-Pf), and Habenular (Hb) in addition to the mammillothalamic tract (MTT).

The two notable differences between HIPS-THOMAS and T1w-THOMAS are the use of cross-correlation metric ("CC" in blue, Fig.1) for ANTs nonlinear registration in calculating the Warp R instead of mutual information metric ("MI" in grey, Fig.1) and the use of joint label fusion instead of majority voting in the final label fusion step.

2. 3 Convolutional Neural Network (CNN) based segmentation

In addition to T1w-THOMAS and HIPS-THOMAS, the CNN method of Umapathy et al. [37] was also used for segmentation of T1w data. This approach uses two cascaded 3D CNNs- WMn-MPRAGE-like images are first synthesized from T1w images using a contrast synthesis CNN and these images are then processed using another CNN to yield thalamic nuclei parcellation. The synthesis network was trained using patches from registered pairs of T1w and WMn-MPRAGE images acquired from the same subjects on 3T GE and Siemens scanners as described in Umapathy et al [37]; the segmentation network was trained using WMn-THOMAS data.

2. 4 Datasets and evaluation metrics

The datasets used in the analysis comprised of 12 subjects acquired on a Siemens 3T scanner with T1w MPRAGE, 19 subjects acquired on a GE 3T scanner with 3D SPGR, and 18 subjects acquired on a Philips 3T scanner with T1w MPRAGE. WMn-MPRAGE acquired on each of these subjects were also available for comparisons. In addition, 8 datasets acquired on a Siemens 7T scanner using Magnetization-Prepared 2 Rapid Acquisition Gradient Echo (MP2RAGE) sequence was also analyzed. All data acquired from human subjects adhered to institutional review board guidelines (University of Arizona for Siemens 3T data; Stanford and SRI International for GE 3T data; Comité de protection des personnes Ile-de-France IV for Philips 3T data; Commission cantonale d'éthique de la recherche sur l'être humain (CER-VD) for Siemens 7T data) and was acquired after prior informed consent in accordance with the Declaration of Helsinki. Additional information about each sequence can be found in the supplementary section. Segmentation performance of T1w-THOMAS, HIPS-THOMAS, and CNN were compared for different-

i. T1w contrast (MPRAGE, SPGR, MP2RAGE)

ii. scanner manufacturers (Siemens, GE, Philips); and

iii. field strengths (3T, 7T).

In the absence of "gold standard" manual segmentations, segmentations were compared to the "silver standard" WMn-THOMAS, using Dice coefficients [42] and percentage volume error for each nucleus. Statistical significance was determined using t-tests with a Bonferroni correction (0.05/13) applied for multiple corrections. Bland-Altman plots were generated to assess the agreement between WMn-THOMAS and T1w-THOMAS or HIPS-THOMAS segmentations on the GE 3T datasets (n=19).

2.5 Clinical validation

To assess clinical validity, results of HIPS-THOMAS and T1w-THOMAS segmentations of T1w data from 21 individuals diagnosed with alcohol use disorders (AUD, 40-74 years old, six women) and 23 gender- and age-matched controls (23-67 years old, eight women) were compared to the analysis using WMn-THOMAS [33]. For each method, an ANCOVA (jamovi 2.3) was performed with age and intracranial volumes as covariates for each thalamic nucleus. To further assess the accuracy of the 3 methods, a ROC analysis was conducted for relevant thalamic nuclei and the area under the ROC curve was computed using a logistic regression. In the ROC analysis, after removing any outliers, volumes were adjusted for age and intracranial volume using the residual method [43]. Acquisitions were from a previously published study [33]. All acquired data from human subjects adhered to institutional review board guidelines (Stanford and SRI International) and was acquired after prior informed consent in accordance with the Declaration of Helsinki.

## 3. Results

### 3.1 HIPS synthesis

The function that maximized the SSIM, VSI, and Dice was a 3$^{rd}$ order polynomial function. A plot of T1w vs. WMn-MPRAGE normalized image intensities (blue dots) is shown in Figure 2a with a 3$^{rd}$ order polynomial fit (green line) for a Philips 3T dataset. The fit curves from 10 different subjects are shown in Figure 2b. To obviate subject-specific curve fitting, the coefficients resulting from the 10 cases were averaged (red dashed line) to form a single polynomial function (Eq. 1).

$$f(x) = 1 + 0.597*x - 2.0067* x^2 + 0.4529* x^3 \quad [Eq.\ 1]$$

The normalization and averaging steps allowed for this same function to be applicable on Siemens 3T, GE 3T, and Siemens 7T data. To evaluate the quality of the synthesis, density plots between native WMn-MPRAGE and WMn-like synthesized images obtained through the application of the averaged function on an example case from each of the 4 scanner types are shown in Figures 2c-f. The linearity of the density plots attests to the quality and robustness of the synthesis, with the Siemens 7T showing the most concordance (minimal dispersion from the unity line).

### 3.2 Qualitative comparisons

Figure 3 shows acquired T1w and WMn-MPRAGE images as well as HIPS and CNN-synthesized WMn-like images for a Siemens 3T subject (a-d) and a GE 3T subject (e-h). The corresponding thalamic nuclei segmentations for the left side are also shown overlaid. Both the CNN and HIPS synthesized WMn images show improved intra-thalamic contrast (brighter signal in MD and pulvinar nuclei) and thalamic boundaries (white arrows) compared to T1w images. The CNN-synthesized images look less noisy with even better contrast compared to HIPS (d,h). In addition to Siemens and GE 3T, T1w and WMn-MPRAGE images as well as HIPS and CNN-synthesized WMn-like images for a Philips 3T subject (i-l) and a Siemens 7T subject (m-p) are also shown in Figure 3. Note that the CNN was trained only using GE and Siemens 3T data as described earlier and the Philips and 7T represent different scan manufacturer and field strength, respectively, as a test of robustness. HIPS-synthesized images look very similar to WMn-MPRAGE images and produce segmentations comparable to those of WMn-THOMAS for both Philips 3T and Siemens 7T subjects. The CNN method failed on more than a third of the Philips cases (Fig. 3i) and on all Siemens 7T cases (Fig. 3p) due to failures in the synthesis step, leading to poor performance of the subsequent segmentation CNN.

### 3. 3 Quantitative assessments

For quantitative comparisons, mean Dice from Siemens 3T MPRAGE (n=12) and GE 3T SPGR (n=19) datasets were compared between T1w-THOMAS, HIPS-THOMAS, and CNN and the % improvement over T1w-THOMAS was computed. HIPS-THOMAS showed significantly improved Dice compared to T1w-THOMAS for 7/11 nuclei and the MTT on both Siemens 3T (Tab. 1) and GE 3T (Tab. 2) data. For HIPS-THOMAS, more than 10% increase in Dice for the VA, VLP, VPL, and LGN nuclei was observed on Siemens 3T MPRAGE and more than 15% increase in Dice for the VA, VLP, VPL, LGN, and CM nuclei was observed on GE 3T SPGR data. By contrast, the CNN performed better on 1/11 nuclei on Siemens 3T data and 6/11 on GE 3T data with decreased Dice for several nuclei (significant only for the MTT). Notably, HIPS-THOMAS outperformed the CNN, with a higher mean and lower standard deviation, for many nuclei, especially the VLP and the MTT for both GE and Siemens data but also the VA, VPL, Pul, MGN, and MD on GE 3T data. Similar to the performance on GE and Siemens 3T, HIPS-THOMAS significantly improved Dice compared to T1w-THOMAS for 8/11 nuclei and the MTT on Philips 3T data and 9/11 nuclei and the MTT on Siemens 7T data (Tab. 3). More than 15% increase in Dice for the VA, VLP, MD nuclei was observed on Philips 3T data and more than 30% increase for the AV, VA, VLa, VLP, VPL, CM, MD nuclei on the Siemens 7T data. Only the Hb nucleus showed decreased Dice using HIPS-THOMAS compared to T1w-THOMAS on SIEMENS 7T (not statistically significant). Subjects who were scanned on 7T were also scanned on 3T and this data is also shown in Table 3. HIPS-THOMAS at 7T is substantially better (>15%) than at 3T on the VLa, LGN, CM nuclei and the MTT.

Mean volume errors (percentage) of T1w-THOMAS, HIPS-THOMAS, and CNN segmentations compared to WMn-THOMAS segmentation are shown in Figure 4 for Siemens 3T (a), GE 3T (b), Philips 3T (c), and Siemens 7T (d) datasets. HIPS-THOMAS had the lowest error in 9 nuclei and the MTT and highest error in 2 nuclei for Siemens 3T data. A similar trend was also observed for the GE 3T data. T1w-THOMAS displayed the highest errors in 8 nuclei on Siemens 3T data and 7 nuclei on GE 3T data. CNN performance was either comparable to or slightly worse than HIPS except in the MTT and the AV nucleus on Siemens data where it was substantially worse (higher error) and the AV nucleus on GE data where it was substantially better (lower error). HIPS-THOMAS had lower mean errors on all nuclei for Philips 3T data and all except the Hb nucleus for Siemens 7T data.

Figure 5 demonstrates the improvements in Dice (%) and reduction in volume error % for HIPS-THOMAS compared to T1w-THOMAS for the 4 scanners (GE 3T, Siemens 3T, Philips 3T, and Siemens 7T). While 7T HIPS-THOMAS displayed the largest increase in Dice and reduction in volume errors, 3T showed at least 15% increase in Dice in ventral nuclei such as the VA and VPL across all scanners. Volume errors were also reduced for HIPS-THOMAS for all the scanners except for the VPl and Hb nuclei on GE 3T data and the Hb nucleus on Siemens 7T data (white regions).

### 3. 4 Bias and reproducibility

Results of the Bland-Altman analysis are shown in Supplemental Figure 1. For all nuclei, HIPS-THOMAS segmentations had lower mean of differences and lower standard deviations compared to T1w-THOMAS. HIPS-THOMAS shows reduced spread and bias even in large nuclei such as the MD, VPL, and VLP nuclei.

### 3. 5 Clinical validation

Results of the ANCOVA analysis on a cohort of healthy subjects and patients with alcohol use disorder are summarized in Table 5. As in Zahr et al. [33], only the ventral thalamic region showed significant atrophy in AUD relative to healthy controls. In this study, the left and right hemispheres were analyzed separately, and atrophy was noted only in the left VLP nucleus in AUD compared to healthy controls after adjusting for age and ICV. While this was significant for WMn-THOMAS (p=0.004), T1w-THOMAS (p=0.011), and HIPS-THOMAS (p=0.008), the p-value for HIPS-THOMAS was much closer to that of WMn-THOMAS than T1w-THOMAS. Furthermore, only T1w-THOMAS also exhibited atrophy in the right VLP nucleus (p=0.003), which was not seen on WMn-THOMAS or HIPS-THOMAS and is likely a spurious result. Using both VLP and VPL nuclei volumes, the area under the ROC curve (AUC) was higher for WMn-THOMAS (0.84) and HIPS-THOMAS (0.79) compared to T1w-THOMAS (0.73).

## 4. Discussion

Here, we have improved the THOMAS pipeline to permit significantly thalamic nuclei segmentation from T1w images, by employing a computationally efficient, polynomial synthesis method to generate WMn-like images prior to segmentation. Robustness of segmentations using HIPS-THOMAS and a single cubic function estimated from averaging 10 Philips 3T datasets was demonstrated from data across differing T1 image contrasts (i.e. SPGR, MPRAGE, MP2RAGE), scanner manufacturers (i.e. Siemens, GE, Philips), and field strengths (i.e. 3T, 7T) as shown by higher Dice indices and lower volume errors compared to T1w-THOMAS and CNN for a majority of the nuclei.

The use of an averaged 3$^{rd}$ order polynomial function on normalized input images in HIPS resulted in synthesized WMn-like images that closely resemble native WMn-MPRAGE images, independent of scanner manufacturer or field strength. At 7T, the MP2RAGE sequence produces significantly higher contrast images than MPRAGE, resulting in synthesized WMn-like images that very closely match the WMn-MPRAGE images as seen by the near perfect linear density plot with minimal spread in Figure 2F. The CNN-synthesized images have less noise and slightly enhanced contrast compared to HIPS images (Fig. 3) due to the inherent denoising in the synthesis CNN. In contrast, the use of polynomial functions in HIPS could result in noise amplification. Preliminary testing of patch based denoising (*DenoiseImage* of ANTs) caused blurring and reduction of contrast in some cases, leading to poorer segmentations. Hence, denoising was not incorporated in the HIPS-THOMAS pipeline but could prove useful in noisy datasets. The significant contribution of HIPS to the THOMAS pipeline is to create images with similar contrast profiles as WMn-MPRAGE images, allowing the use of CC metric for nonlinear registration to the WMn template, which is more accurate than the MI metric used in T1w-THOMAS [44]. The joint fusion algorithm used in HIPS-THOMAS (cf. majority voting in T1w-THOMAS), also likely contributed to increased label accuracy [35] [36]. HIPS is computationally efficient, does not add much complexity to the image analysis pipeline of THOMAS, and does not require separate training for the different scenarios of scanner manufacturers and field strengths, as required by CNN-based methods.

Segmentations of T1w data using HIPS-THOMAS are closer than T1w-THOMAS to THOMAS segmentations using WMn-MPRAGE images, resulting in improved volume accuracy and Dice for a majority of the nuclei. Dice was most improved for ventral nuclei (VA, VLP, VPL) where the improved contrast in T1w images better delineates external boundaries. The HIPS-THOMAS Dice improvement reaches its highest performance at 7T (Tab. 3), likely due to the more accurate synthesis of WMn-like images at 7T using MP2RAGE T1 maps. Only the Hb nucleus has a decreased Dice index using HIPS-THOMAS on Siemens 7T data compared to the use of T1w-THOMAS (although not statistically significant), which may be explained by a higher variability of segmentations for this nucleus and a small effective in this dataset. Quantitative robustness of HIPS is further demonstrated by the Bland-Altman plots (Supplementary Fig. 1) with segmentations from HIPS-THOMAS displaying smaller bias and less dispersion compared to T1w-THOMAS. Lastly, HIPS-THOMAS was validated in an AUD cohort, demonstrating consistent results with WMn-THOMAS, namely reduced the VLp nucleus volumes in AUD patients relative to healthy controls. In contrast, T1w-THOMAS showed much worse sensitivity for the left VPL nucleus (p 0.011) and a potentially spurious atrophy of the right VPL nucleus. These results are analogous to the results of Umapathy et al. [37] where the synthesis-segmentation CNN displayed better accuracy compared to a segmentation network trained directly on T1w images, which showed spurious VPL atrophy.

CNN performs comparably to HIPS-THOMAS on sequences on which it was trained (i.e. Siemens 3T MPRAGE) or on similar contrast images (SPGR), but HIPS-THOMAS is considerably better than the CNN for Philips 3T and Siemens 7T data, where the CNN failed on a majority of cases due to lack of adequate training. While HIPS-THOMAS can yield higher Dice scores, it may paradoxically have higher volume errors than the CNN for certain nuclei on Siemens 3T (e.g. VPL) and GE 3T data (e.g. AV, LGN), as illustrated in Figure 4. This discrepancy may be due to inherent denoising in CNNs, which allows for more precise delimitation. In summary, HIPS-THOMAS is more flexible and generalizable, as it can be applied easily to T1w images from different scanners without requiring training, a big advantage considering public databases like ADNI and OASIS contain data from a mix of manufacturers and field strengths.

Our work had some limitations. The T1w segmentations were evaluated against segmentations from THOMAS applied to WMn-MPRAGE images, which is not ideal. THOMAS has been thoroughly validated against manual segmentation at 3T (Su et al, ISMRM abstract) [45] and 7T [32] and was adopted as a "silver" standard, substituting for the "gold" standard manual segmentation, which is very time consuming and requires specific domain expertise. The performance of the CNN method could also be enhanced by training the synthesis and segmentation networks with data from Philips and 7T scanners but was beyond the scope of this work. In the future, more complex exemplar-based synthesis approaches such as MIMECS (MR image example-based contrast synthesis) [46], could replace the simpler 3$^{rd}$ order polynomial approach taken here, which could further improve HIPS performance and reduce noise amplification. This could also be useful for pipelines for analyzing whole brain images (as opposed to cropped images like in our current pipeline) where artifacts from scalp and other sources can impair HIPS performance. HIPS showed Dice improvements in several nuclei, but the Hb and VLa nuclei consistently showed < 0.7 Dice across all 3T scanners. While the poor habenula Dice could be attributed to its small size, the reasons for suboptimal VLa Dice compared to other ventral nuclei like the VA warrants further investigation.

## 5. Conclusion

WMn-like images synthesized using the computationally efficient HIPS significantly improved the robustness as well as the accuracy of THOMAS compared to direct THOMAS or CNN-based methods for segmentation of T1w data.

## 6. Data availability statement

The datasets analysed during the current study are available from the corresponding author on reasonable request.

## 7. Code availability

The code used in the current study is available at https://github.com/thalamicseg/hipsthomasdocker.

## 9. Acknowledgements

We wish to acknowledge Germain Arribarat for his support and his time during the beginning of this project, forming J.P.V to image processing. This work was supported by the University of Toulouse 3 Paul Sabatier via a PhD student mobility grant obtained by J.P.V.

## 10. Author contributions

J.P.V. and M.S. conceived and developed the presented method. J.P.V. performed the computations and analysis under the supervision of M.S. J.P.V. wrote the manuscript with support from L.D., E.J.B., and mainly with M.S.

L.D., E.J.B., P.P., J.P., M.B.C., N.M.Z. acquired and shared MRI data from their previous respective studies and helped with the study design.

## 11. Additional information

Competing interests statement: The authors have no conflicts of interest relevant to this manuscript to disclose.

# Figure legends and tables

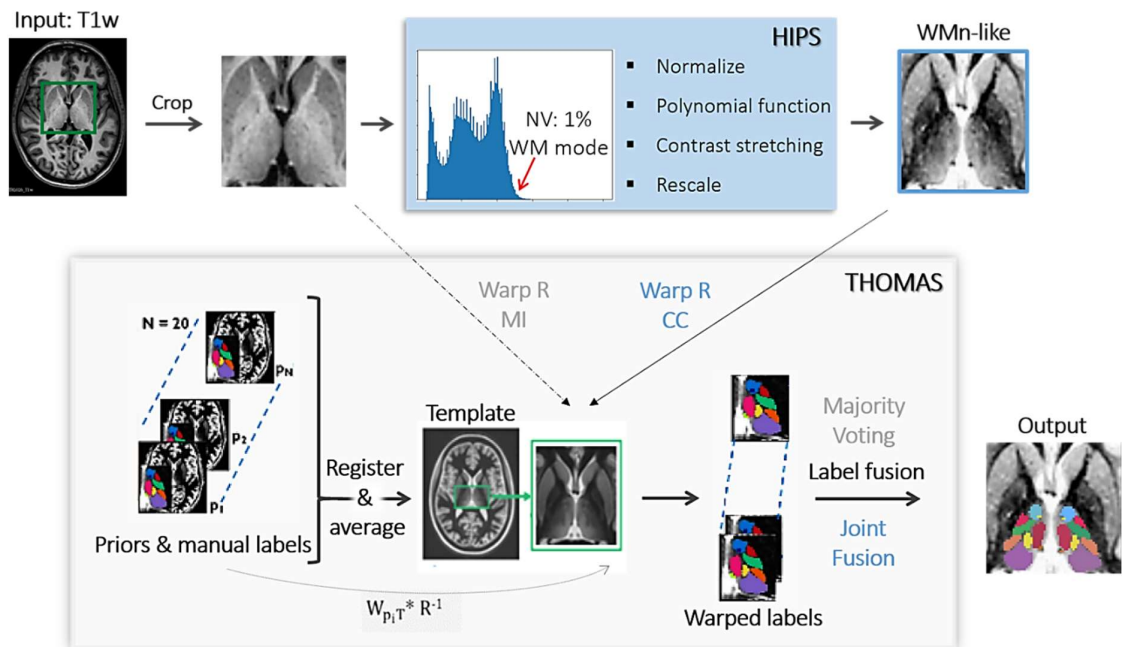

Figure 1: Proposed HIPS-THOMAS pipeline. HIPS pre-processing includes normalization of the cropped T1w input, application of a polynomial function to generate a WMn-like image, and a contrast stretching step followed by a rescaling step. The WMn-like cropped input is fed into the THOMAS pipeline as opposed to the original cropped T1w image. Note that for HIPS-THOMAS, the nonlinear warp R uses a cross-correlation metric (CC in blue) and the label fusion step uses joint label fusion (in blue) as opposed to mutual information metric (MI in grey) and majority voting (in grey).

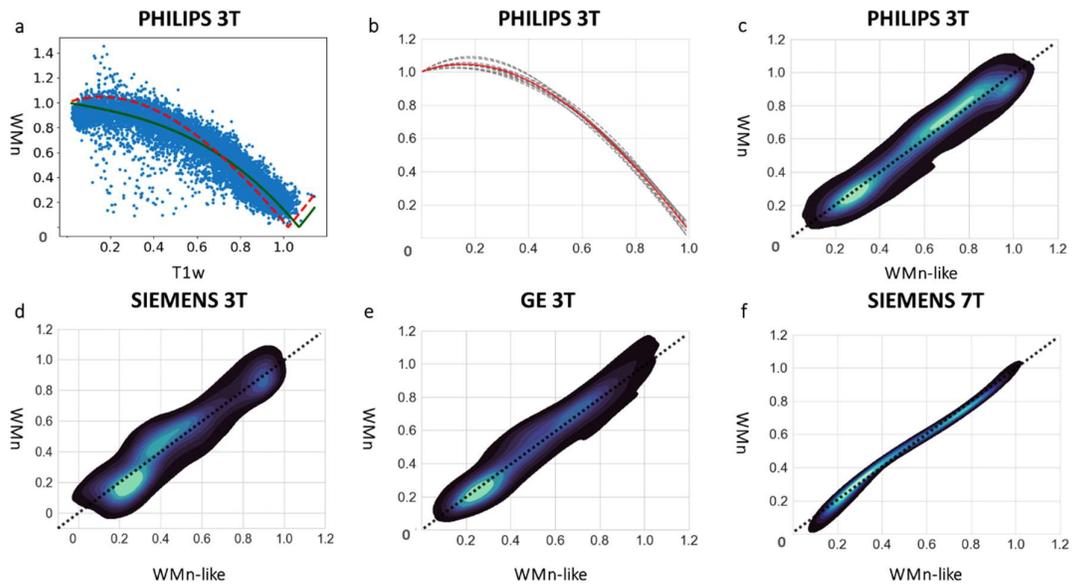

Figure 2: HIPS construction and performance. (a) Normalized intensity plot between WMn-MPRAGE and T1-MPRAGE data from a Philips 3T subject (blue dots) and a 3rd order polynomial fit (green line). (b) The 10 curves resulting from the individual fitting on 10 Philips data cases (dashed gray) and the resulting averaged function (red line in a, b). Density plots between normalized WMn-MPRAGE and synthesized WMn-MPRAGE data using the same averaged function on an example subject from Philips 3T (c), Siemens 3T (d), GE 3T (e), and Siemens 7T (e). The black dashed unity line represents perfect concordance between images.

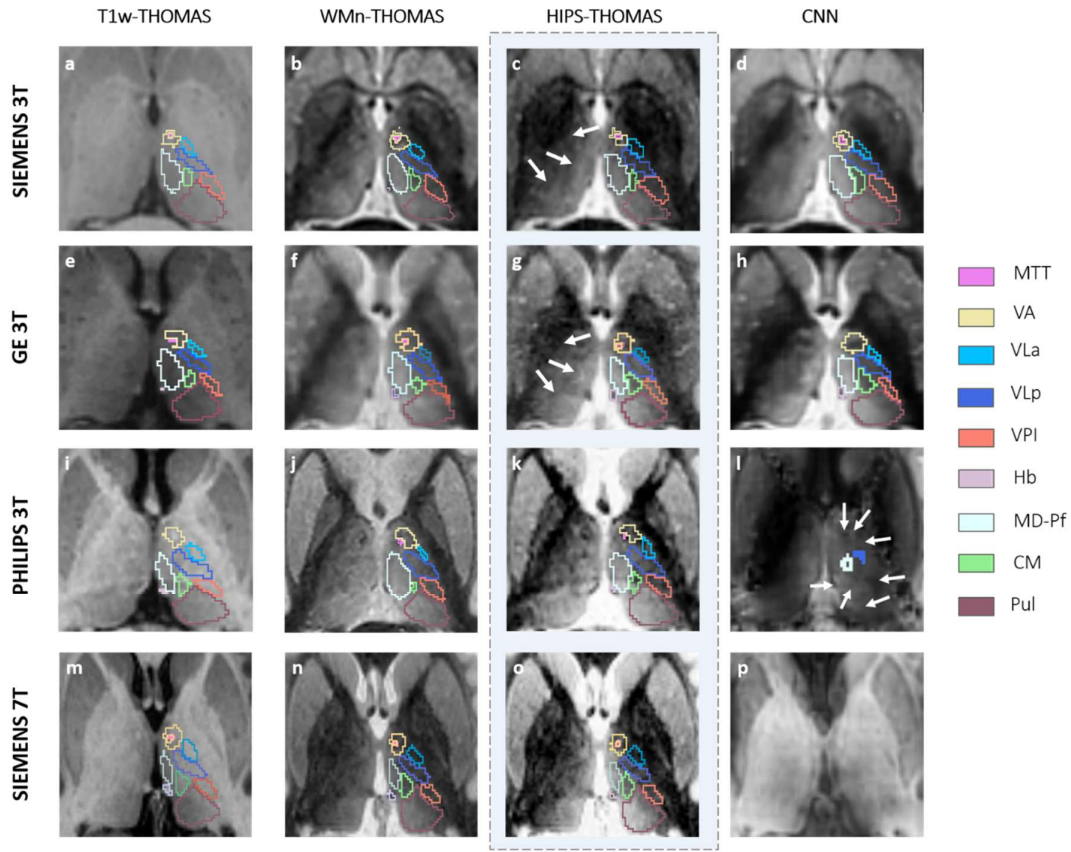

Figure 3: An axial slice from acquired T1w and WMn-MPRAGE as well as HIPS and CNN-synthesized WMn images for a Siemens 3T MPRAGE (a-d), GE 3T SPGR (e-h), Philips 3T MPRAGE (i-l) or Siemens 7T MP2RAGE (m-p) subject with the corresponding nuclei segmentations overlaid on the left thalamus. Note the improved intrathalamic contrast and thalamic boundaries (white arrows on c and g) in the synthesized WMn-like images produced by HIPS-THOMAS compared to the native T1w images. The failure of CNN synthesis can clearly be seen panels i and p along with the failed or missing (white arrows in l) segmentations.

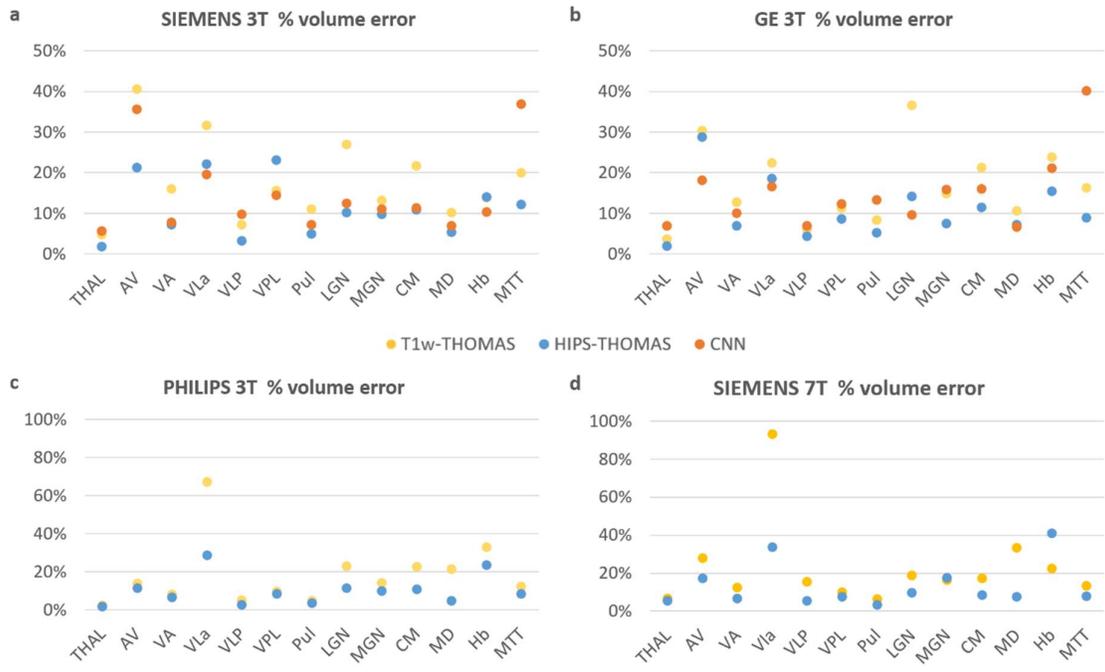

Figure 4: Mean volume error (%) of T1w-THOMAS, HIPS-THOMAS, and CNN segmentations, compared to WMn-THOMAS segmentations for Siemens 3T MPRAGE (a, n=12) and GE 3T SPGR (b, n=19) Philips 3T MPRAGE 3T (c, n=18) and Siemens 7T MP2RAGE (d, n=8) data. The general trend of HIPS < CNN < T1w THOMAS was observed for most nuclei at 3T. HIPS-THOMAS errors were lower than T1w-THOMAS for all nuclei but Hb for Siemens 7T data.

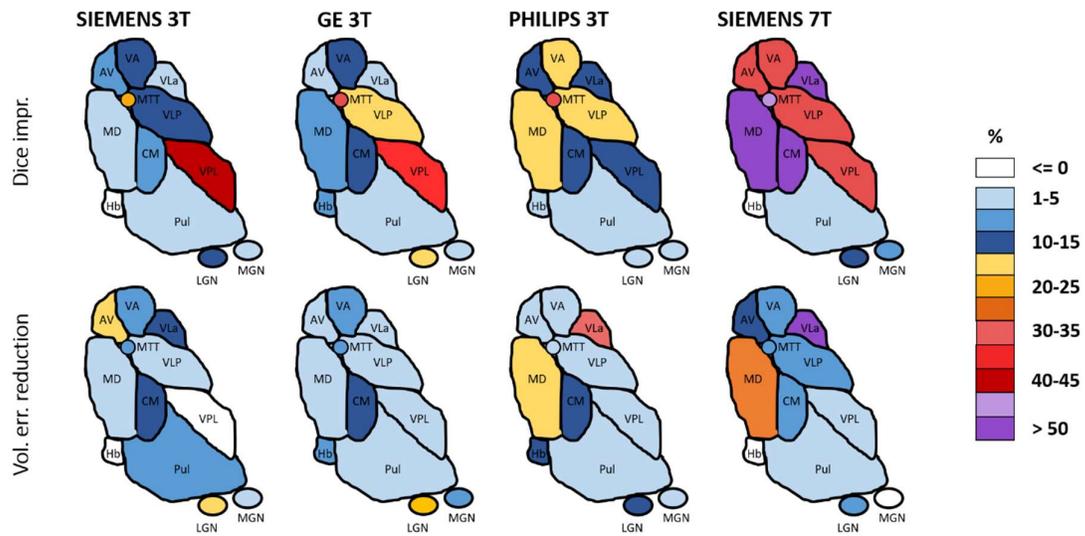

Figure 5: Graphical summary of improvement of mean Dice (%) and reduction of % volume error for each thalamic nucleus on Siemens 3T, GE 3T, PHILIPS 3T and Siemens 7T data using HIPS-THOMAS compared to T1w-THOMAS. Labels: cf. Method 2.2.

|  | **T1w-THOMAS** | **SIEMENS 3T** HIPS-THOMAS | Impr. | CNN | Impr. |
|---|---|---|---|---|---|
| **THAL** | 0.91 +/- 0.01 | *** 0.93 +/- 0.01 | 2% | 0.91 +/- 0.04 | 1% |
| **AV** | 0.67 +/- 0.10 | ** 0.73 +/- 0.07 | 7% | 0.71 +/- 0.18 | 6% |
| **VA** | 0.69 +/- 0.03 | ** 0.77 +/- 0.05 | 12% | 0.72 +/- 0.09 | 3% |
| **VLa** | 0.61 +/- 0.06 | 0.64 +/- 0.07 | 5% | 0.61 +/- 0.12 | 0% |
| **VLP** | 0.76 +/- 0.04 | *** 0.86 +/- 0.02 | 13% | 0.81 +/- 0.06 | 6% |
| **VPL** | 0.52 +/- 0.15 | ** 0.76 +/- 0.03 | 46% | * 0.70 +/- 0.11 | 33% |
| **Pul** | 0.85 +/- 0.03 | *** 0.88 +/- 0.02 | 4% | 0.86 +/- 0.05 | 2% |
| **LGN** | 0.69 +/- 0.08 | *** 0.77 +/- 0.05 | 12% | 0.67 +/- 0.17 | -2% |
| **MGN** | 0.76 +/- 0.04 | 0.78 +/- 0.03 | 3% | 0.70 +/- 0.20 | -8% |
| **CM** | 0.72 +/- 0.05 | 0.76 +/- 0.03 | 6% | 0.71 +/- 0.14 | -2% |
| **MD-Pf** | 0.83 +/- 0.05 | * 0.88 +/- 0.03 | 6% | 0.85 +/- 0.08 | 3% |
| **Hb** | 0.66 +/- 0.04 | 0.66 +/- 0.06 | 0% | 0.60 +/- 0.20 | -9% |
| **MTT** | 0.50 +/- 0.08 | * 0.63 +/- 0.05 | 24% | *** 0.29 +/- 0.11 | -42% |

Table 1: Mean Dice +/- SD of T1w-THOMAS, HIPS-THOMAS, and CNN segmentations along with % improvement for HIPS-THOMAS and CNN compared to T1w-THOMAS for 3T Siemens MPRAGE (n=12) data. HIPS-THOMAS improves Dice by >10% in 4 nuclei and MTT compared to 1 nucleus for CNN. Colors: > 10% green, >0% yellow, <0% red. P-values are corrected for multiple comparisons (Bonferonni): *<0.00385 **<0.001 ***<0.0001.

|  | | GE 3T | | | |
|---|---|---|---|---|---|
|  | **T1w-THOMAS** | **HIPS-THOMAS** | Impr. | **CNN** | Impr. |
| **THAL** | 0.91 +/- 0.01 | *** 0.93 +/- 0.01 | 2% | 0.92 +/- 0.01 | 1% |
| **AV** | 0.68 +/- 0.09 | 0.69 +/- 0.12 | 2% | 0.70 +/- 0.08 | 4% |
| **VA** | 0.70 +/- 0.06 | *** 0.81 +/- 0.02 | 15% | * 0.76 +/- 0.03 | 12% |
| **VLa** | 0.64 +/- 0.08 | 0.66 +/- 0.06 | 4% | 0.62 +/- 0.08 | -2% |
| **VLP** | 0.74 +/- 0.08 | *** 0.85 +/- 0.03 | 16% | *** 0.81 +/- 0.04 | 12% |
| **VPL** | 0.56 +/- 0.14 | *** 0.78 +/- 0.06 | 40% | *** 0.72 +/- 0.07 | 40% |
| **Pul** | 0.85 +/- 0.03 | *** 0.89 +/- 0.02 | 5% | 0.87 +/- 0.02 | 2% |
| **LGN** | 0.67 +/- 0.04 | *** 0.78 +/- 0.04 | 17% | *** 0.75 +/- 0.04 | 13% |
| **MGN** | 0.78 +/- 0.04 | 0.82 +/- 0.04 | 5% | 0.69 +/- 0.10 | -9% |
| **CM** | 0.67 +/- 0.09 | *** 0.77 +/- 0.06 | 15% | * 0.75 +/- 0.03 | 11% |
| **MD-Pf** | 0.81 +/- 0.05 | *** 0.88 +/- 0.02 | 8% | * 0.86 +/- 0.02 | 6% |
| **Hb** | 0.66 +/- 0.07 | 0.70 +/- 0.04 | 6% | 0.67 +/- 0.12 | 3% |
| **MTT** | 0.47 +/- 0.10 | *** 0.63 +/- 0.05 | 32% | ** 0.31 +/- 0.08 | -34% |

Table 2: Mean Dice +/- SD of T1w-THOMAS, HIPS-THOMAS, and CNN segmentations along with % improvement for HIPS-THOMAS and CNN compared to T1w-THOMAS for GE 3T SPGR data (n=19). HIPS-THOMAS improves Dice by >15% in 5 nuclei and MTT compared to 1 nucleus for CNN. Colors: > 10% green, >0% yellow, <0% red. P-values are corrected for multiple comparisons (Bonferonni): *<0.00385 **<0.001 ***<0.0001.

|  | PHILIPS 3T | | | SIEMENS 7T | | | SIEMENS 3T |
|---|---|---|---|---|---|---|---|
|  | T1w-THOMAS | HIPS-THOMAS | Impr. | T1w-THOMAS | HIPS-THOMAS | Impr. | HIPS-THOMAS |
| THAL | 0.92 +/- 0.01 | *** 0.94 +/- 0.00 🟡 | 2% | 0.92 +/- 0.02 | ** 0.96 +/- 0.02 🟡 | 5% | *** 0.93 +/- 0.01 |
| AV | 0.72 +/- 0.06 | *** 0.80 +/- 0.03 🟢 | 11% | 0.64 +/- 0.16 | * 0.84 +/- 0.08 🟢 | 32% | 0.73 +/- 0.10 |
| VA | 0.71 +/- 0.07 | *** 0.82 +/- 0.02 🟢 | 16% | 0.63 +/- 0.15 | * 0.85 +/- 0.06 🟢 | 35% | 0.80 +/- 0.05 |
| VLa | 0.58 +/- 0.12 | 0.64 +/- 0.07 🟡 | 10% | 0.48 +/- 0.12 | ** 0.81 +/- 0.14 🟢 | 69% | * 0.52 +/- 0.15 |
| VLP | 0.76 +/- 0.19 | 0.88 +/- 0.01 🟢 | 16% | 0.70 +/- 0.10 | ** 0.92 +/- 0.04 🟢 | 32% | * 0.87 +/- 0.02 |
| VPL | 0.73 +/- 0.07 | *** 0.81 +/- 0.02 🟢 | 11% | 0.68 +/- 0.10 | ** 0.90 +/- 0.05 🟢 | 33% | * 0.80 +/- 0.03 |
| Pul | 0.87 +/- 0.02 | ** 0.89 +/- 0.02 🟡 | 2% | 0.88 +/- 0.03 | * 0.93 +/- 0.03 🟡 | 5% | * 0.88 +/- 0.02 |
| LGN | 0.74 +/- 0.04 | *** 0.79 +/- 0.03 🟡 | 8% | 0.77 +/- 0.05 | ** 0.88 +/- 0.06 🟢 | 14% | *** 0.76 +/- 0.02 |
| MGN | 0.77 +/- 0.04 | ** 0.80 +/- 0.03 🟡 | 3% | 0.79 +/- 0.08 | 0.86 +/- 0.08 🟡 | 9% | 0.75 +/- 0.05 |
| CM | 0.70 +/- 0.08 | *** 0.79 +/- 0.04 🟢 | 13% | 0.51 +/- 0.20 | ** 0.88 +/- 0.06 🟢 | 73% | * 0.74 +/- 0.08 |
| MD-Pf | 0.76 +/- 0.07 | *** 0.89 +/- 0.01 🟢 | 17% | 0.59 +/- 0.20 | * 0.92 +/- 0.03 🟢 | 58% | ** 0.86 +/- 0.03 |
| Hb | 0.58 +/- 0.08 | 0.59 +/- 0.04 🟡 | 1% | 0.69 +/- 0.12 | 0.63 +/- 0.15 🔴 | -10% | 0.64 +/- 0.07 |
| MTT | 0.48 +/- 0.08 | *** 0.66 +/- 0.03 🟢 | 38% | 0.54 +/- 0.16 | * 0.78 +/- 0.10 🟢 | 46% | ** 0.58 +/- 0.07 |

Table 3: Mean Dice +/- SD of T1w-THOMAS and HIPS-THOMAS segmentations along with % improvement for HIPS-THOMAS compared to T1w-THOMAS for Philips 3T MPRAGE (n=18) and Siemens 7T MP2RAGE (n=8) data. HIPS-THOMAS improves Dice by >10% in 6 nuclei and MTT on Philips 3T data and by >30% in 7 nuclei and MTT on Siemens 7T data. The rightmost column shows Dice results for Siemens 3T MPRAGE data obtained on the same 8 subjects scanned on Siemens 7T. Colors: > 10% green, >0% yellow, <0% red. P-values are corrected for multiple comparisons (Bonferonni): *<0.00385 **<0.001 ***<0.0001.

# Supplemental data

## MRI Datasets

- Philips **T1w** MPRAGE were acquired on a 3T scanner (Achieva dStream, PHILIPS) with the following parameters: 0.9*0.9*1mm voxel size, TR/TE=8.2/3.7, flip angle= 8°, 256x256 matrix, slice thickness= 1 mm, slice number= 170. **WMn-MPRAGE** MRI parameters: 0.68*0.68*0.7 mm voxel size, TR/TE=9.7/4.7, flip angle= 8°, 336x336 matrix, slice thickness= 1.4 mm, slice number= 120. (This is unpublished data from Danet, L., Péran, P., Pariente, J., and Barbeau, E.)

- GE **T1w** SPGR were acquired on a 3T scanner (MR750, General Electric Healthcare) with the following parameters [1]: voxel size, TR/TE/TI= 6.008/1.952/300, flip angle= 9°, NA matrix, slice thickness= 1.3 mm, slice number= 120. **WMn-MPRAGE** scans were collected with parameters: TR/TE/TI= 11/5/500, flip angle= 7, 200x200 matrix, slice thickness= 1 mm, slice number=210.

- Siemens **T1w** MPRAGE were acquired on a scanner at 3T (Magnetom Prisma fit, Siemens Medical Solutions) with the following parameters: voxel size, TR/TE= 2000/2.52, flip angle= 12, 256x256 matrix, slice thickness= 1 mm, slice number= 192. **WMn-MPRAGE** scans were collected with parameters: TR/TE/TI= 11/5/450, flip angle= 7, 224x224 matrix, slice thickness= 1.1 mm, slice number= 160. (This is unpublished data from Saranathan, M.)

- Siemens **T1w** MP2RAGE were acquired on a scanner at 7T (Magnetom, Siemens Medical Solutions) with the following parameters [2]: 0.8*0.8*0.8 or 0.6*0.6*0.6 voxel size, TR/TE/TI=6000/2.64 or 2.05/800 or 2700, flip angle= 7 or 5°, 240x256 or 256x320 matrix, slice thickness= NA, slice number= 320. **WMn-MPRAGE** images were synthesized using the inversion recovery equation (1- 2 exp(-TI/$T_1$) with the inversion time TI=670ms and $T_1$ being the $T_1$ maps produced from MP2RAGE acquisitions.

- For the clinical validation AUD dataset details, refer to Zahr et al [1].

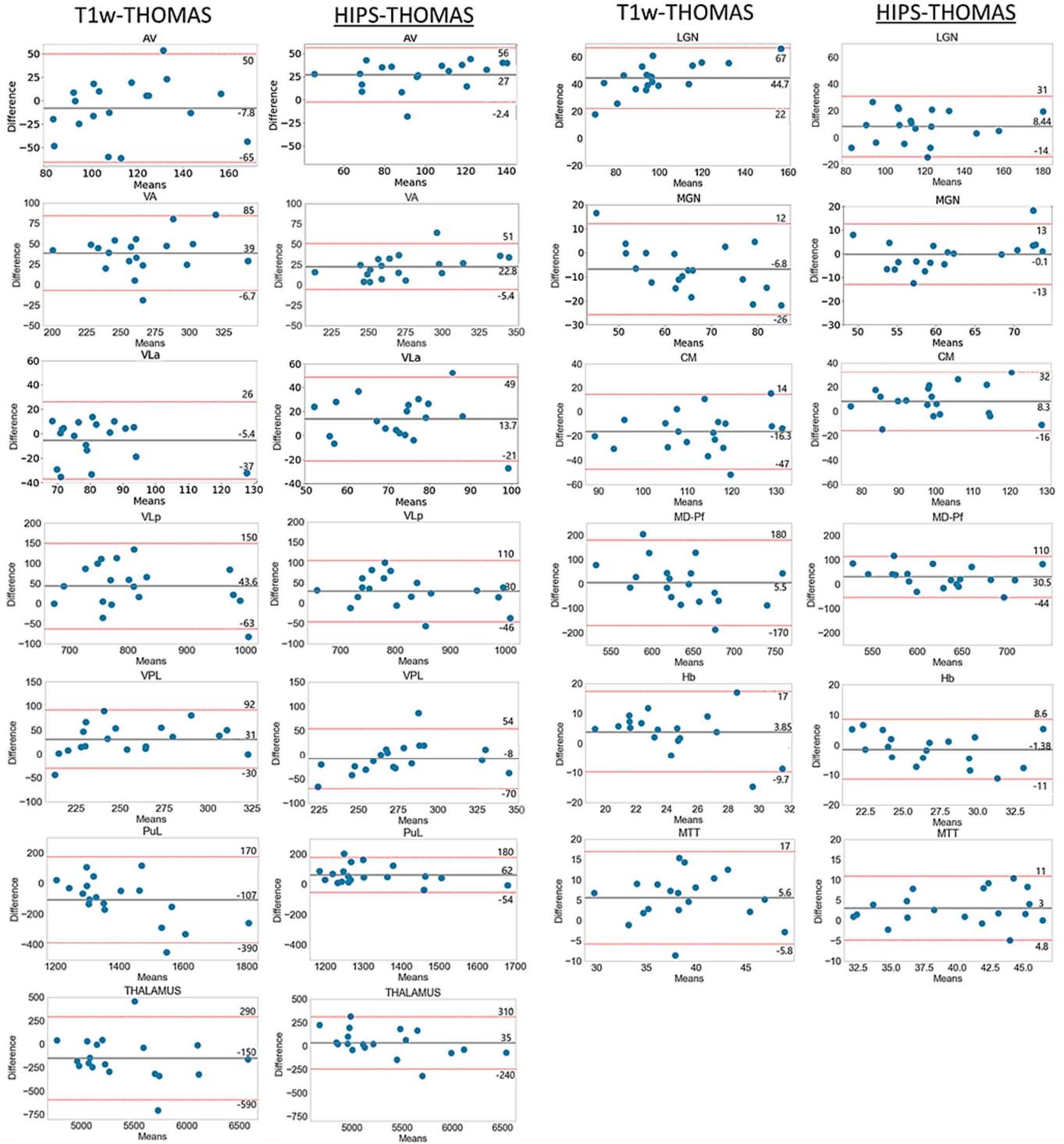

Supplementary Figure 1: Bland-Altman plots between WMn-THOMAS vs. T1w-THOMAS and HIPS-THOMAS for the whole left thalamus and left nuclei volumes of GE 3T SPGR dataset (n=19). The black line represents the mean of the difference between the two methods and red lines are +/- 1.96 SD with their corresponding values.